\documentclass[%
 aps,
prl,%
 amsmath,amssymb,
reprint,%
superscriptaddress]{revtex4-1}

\usepackage{graphicx}% Include figure files
\usepackage{dcolumn}% Align table columns on decimal point
\usepackage{bm}% bold math
\usepackage{color}
\newcommand{\comment}[1]{}
\begin{document}

%\preprint{AIP/123-QED}

\title[Role of phonon skew scattering in the spin Hall effect of platinum]{Role of phonon skew scattering in the spin Hall effect of platinum}
\author{G. V. Karnad}
\affiliation{Institut f{\"u}r Physik, Johannes Gutenberg-Universit{\"a}t, Staudinger Weg 7, 55128 Mainz, Germany}
\author{C. Gorini}
\email{cosimo.gorini@physik.uni-regensburg.de}
\affiliation{Institut f{\"u}r Theoretische Physik, Universit{\"a}t Regensburg, 93040 Regensburg, Germany}

\author{K. Lee}
\affiliation{Institut f{\"u}r Physik, Johannes Gutenberg-Universit{\"a}t, Staudinger Weg 7, 55128 Mainz, Germany}
\author{T. Schulz}
\affiliation{Institut f{\"u}r Physik, Johannes Gutenberg-Universit{\"a}t, Staudinger Weg 7, 55128 Mainz, Germany}

\author{R. Lo Conte}
\affiliation{Institut f{\"u}r Physik, Johannes Gutenberg-Universit{\"a}t, Staudinger Weg 7, 55128 Mainz, Germany}
\affiliation{Graduate School of Excellence \textquotedblleft{Materials Science in Mainz}\textquotedblright (MAINZ), Staudinger Weg 9, 55128 Mainz, Germany}
\author{A. W. J. Wells}
\affiliation{School of Physics and Astronomy, University of Leeds, Leeds LS2 9JT, United Kingdom}
\author{D.-S. Han}
\affiliation{Department of Applied Physics, Eindhoven University of Technology, 5600 MB Eindhoven, The Netherlands}
\author{K. Shahbazi}
\affiliation{School of Physics and Astronomy, University of Leeds, Leeds LS2 9JT, United Kingdom}
\author{J.-S. Kim}
\altaffiliation[now at: ]{DGIST Research Center for Emerging Materials, DIGST, Daegu, 42988, Republic of Korea}
\affiliation{Department of Applied Physics, Eindhoven University of Technology, 5600 MB Eindhoven, The Netherlands}

%\altaffiliation[optional text]{affiliation information}
\author{T. A. Moore}
\affiliation{School of Physics and Astronomy, University of Leeds, Leeds LS2 9JT, United Kingdom}
\author{H. J. M. Swagten}
\affiliation{Department of Applied Physics, Eindhoven University of Technology, 5600 MB Eindhoven, The Netherlands}
\author{U. Eckern}
\affiliation{Institut f{\"u}r Physik, Universit{\"a}t Augsburg, 86135 Augsburg, Germany}
\author{R. Raimondi}
\affiliation{Dipartimento di Matematica e Fisica, Roma Tre University, Via della Vasca Navale 84, 00146 Rome, Italy}
\author{M. Kl{\"a}ui}
\email{klaeui@uni-mainz.de}
\affiliation{Institut f{\"u}r Physik, Johannes Gutenberg-Universit{\"a}t, Staudinger Weg 7, 55128 Mainz, Germany}
\affiliation{Graduate School of Excellence \textquotedblleft{Materials Science in Mainz}\textquotedblright (MAINZ), Staudinger Weg 9, 55128 Mainz, Germany}

\date{\today}% It is always \today, today,

\begin{abstract}
We measure and analyze the effective spin Hall angle of platinum in the low residual resistivity  regime by second harmonic measurements of the spin-orbit torques for a multilayer of Pt$\mid$Co$\mid$AlO$_{x}$. An angular dependent study of the torques allows us to extract the  effective spin Hall angle responsible for the damping-like torque in the system. We observe a strikingly non-monotonic and reproducible  temperature dependence of the torques. This behavior is compatible with recent theoretical predictions which include both intrinsic and extrinsic (impurities and phonons) contributions to the spin Hall effect at finite temperature.

%Valid PACS numbers may be entered using the \verb+\pacs{#1}+ command.
\end{abstract}

\maketitle
The demand for high density and efficient data storage devices has driven the research on magnetic memories. Proposals were mainly based on using spin transfer torques for domain wall motion and magnetization switching \cite{Prenat2015}. However, spin transfer torque-based devices are plagued by the requirement of high writing current densities which result in deterioration of the material layers and hence limit the device life. 
Recent demonstration \cite{liu_spin-torque_2012} of highly efficient current-induced switching via spin orbit torques (SOTs) shows promising results that can overcome these limitations \cite{Prenat2015}. Such devices are made from magnetic multi-layers which consist of a few mono-layers thick ferromagnetic (FM) material sandwiched between a normal metal (NM) and an oxide layer.

The injection of an in-plane current through the normal metal has been shown to efficiently manipulate the magnetization \cite{Liu_2012_PRL}. The torques have been attributed to the spin Hall effect (SHE) \cite{Liu_2012_PRL,garello_symmetry_2013,guo_intrinsic_2008,Sinova_2015} arising from the charge to spin current conversion in the NM and the inverse spin galvanic effect (ISGE) \cite{aronov1989nuclear, edelstein_spin_1990,ganichev_spin-galvanic_2002, shen_microscopic_2014,borge_spin_2014,miron2011perpendicular,ganichev_spin_2016} arising at the interfaces with the FM.

\begin{figure}
%\centering
\includegraphics[width = 1\linewidth]{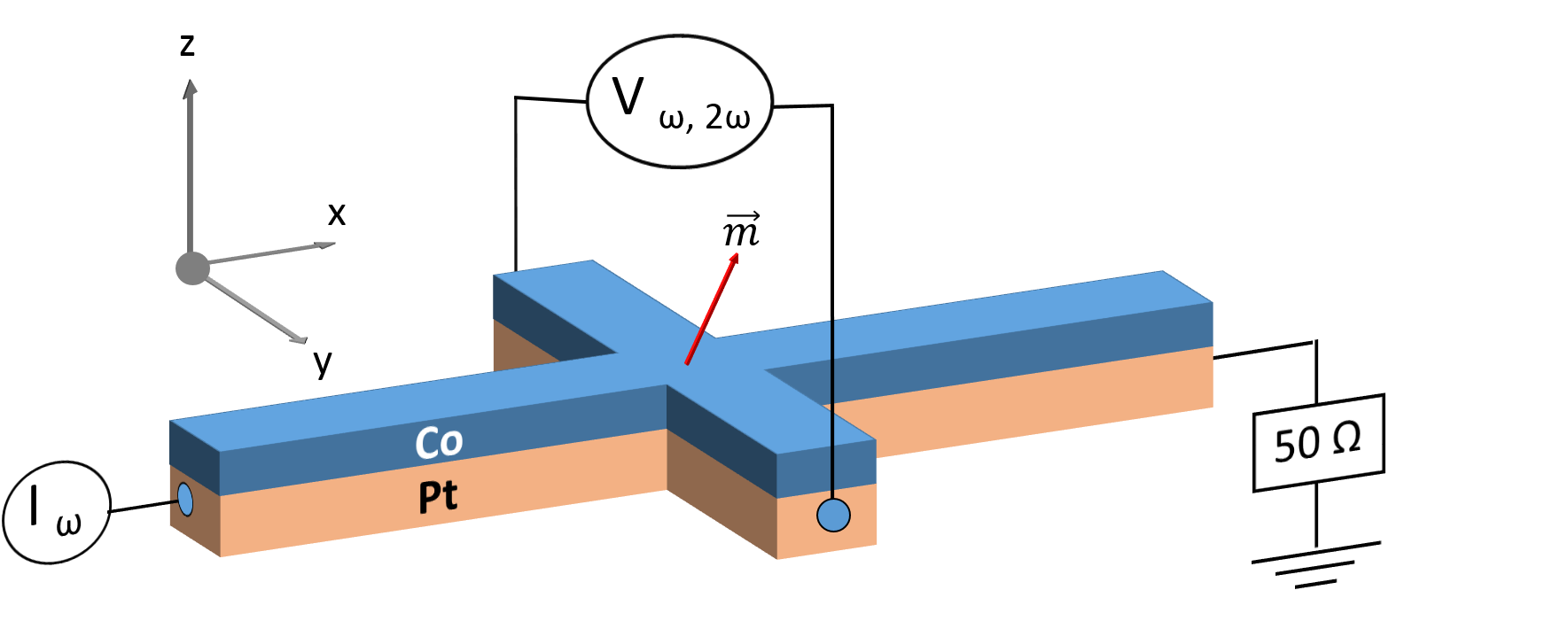}
\caption{Schematics of the second harmonic measurement setup.  A charge current is injected along $\hat{x}$, generating
damping-like and field-like effective spin-orbit fields which torque the magnetization $\hat{m}$.  
The damping-like torque acts along $\hat{\sigma}\times\hat{m}$, the field-like one along $\hat{\sigma}$.
Here $\hat{\sigma}$ is the direction of the current-induced non-equilibrium spin current (SHE) and density (ISGE).} 
\label{fig:SetupSchematic}
\end{figure}

Firstly, a charge current in the NM underlayer generates via the SHE a spin current flowing towards the FM layer.  
These spins enter the FM and exert by $s$-$d$ exchange a damping-like torque on the FM magnetization
\cite{liu_spin-torque_2012,garello_symmetry_2013,hayashi_quantitative_2014,Liu_2012_PRL}.  
Secondly, due to inversion symmetry breaking at the NM$\mid$FM interface, the same charge current gives also rise to an (interfacial) 
non-equilibrium spin polarization via the ISGE \cite{shen_microscopic_2014,borge_spin_2014,wang2012diffusive}.  
Such spin polarization results in a field-like torque affecting the FM magnetization.  
It has actually been discussed how the SHE and the ISGE can {\it individually} in principle generate both damping-like and field-like torques, due to subtler interfacial spin-orbit effects \cite{haney_current_2013, AminPheno_2016}.  
Indeed, we will argue later that in our system the SHE is dominating and responsible for both torques.  
Irrespective of their origin, whether SHE or ISGE, each torque can be described in terms of the corresponding component of an overall,
current-induced spin-orbit field acting on the magnetization: a longitudinal component $\mu_0 H_{DL} \sim \hat{\sigma} \times \hat{m}$,
exerting a damping-like torque $\sim\hat{m} \times (\hat{\sigma} \times \hat{m})$, 
and a transversal one $\mu_0 H_{FL} \sim \hat{\sigma}$, giving rise to a field-like torque $\sim \hat{m} \times \hat{\sigma}$.  
Here $\hat{m}$ is the magnetization direction, and $\hat{\sigma}$ that of the spin of the (non-equilibrium) SHE and/or ISGE electrons.  
In our experimental configuration the charge current flows along $\hat{x}$, while $\hat{m} = \hat{z}, \hat{\sigma} = \hat{y}$, 
see Fig.~\ref{fig:SetupSchematic}.

Efficient magnetization switching using SOTs has motivated studies of various heavy metals, so as to find materials 
with the highest spin Hall angle (SHA).  The SHA characterizes the efficiency of charge-to-spin conversion, 
and is defined as $\theta_{sH} = \sigma_{sH}/\sigma = -\rho_{sH}/\rho$, with $\sigma (\rho)$ and $\sigma_{sH} (\rho_{sH})$ 
respectively the charge and spin Hall conductivities (resistivities).
There are various mechanisms giving rise to the SHE: extrinsic (side-jump \cite{berger_side-jump_1970} 
and skew scattering \cite{smit_spontaneous_1958,dyakonov1971}) and Berry curvature-induced intrinsic ones \cite{sinova_universal_2004}.  
In metallic paramagnets it is difficult to identify materials where only one mechanism is at work.  
It is important to rather understand the dominant mechanism in a particular system in order to optimize it for possible technological
applications.  Indeed, phonon skew scattering has been recently suggested to play an important role 
at room temperature \cite{gorini_spin_2015}, and a thorough experimental study of its influence on the SHE is still lacking.

In this Letter we provide such a study, by monitoring the temperature dependence of the SOTs ascribed to the SHE in low-residual 
resistivity Pt (5.3 $\mu \Omega$cm, 6.3 $\mu \Omega$cm).  While the observed non-monotonic behavior is unexpected and cannot be explained following the established phenomenological analysis \cite{Tian_2009,isasa_temperature_2015}, it is compatible with the temperature scaling derived from the microscopic theory \cite{gorini_spin_2015}. The importance of ``ultra-clean" samples is further supported by the simpler monotonic behavior observed in high-residual resistivity Ta.

%\begin{figure}[!ht]

\begin{figure}[!ht]
%\begin{figure}
\centering
\includegraphics[width = 1\linewidth]{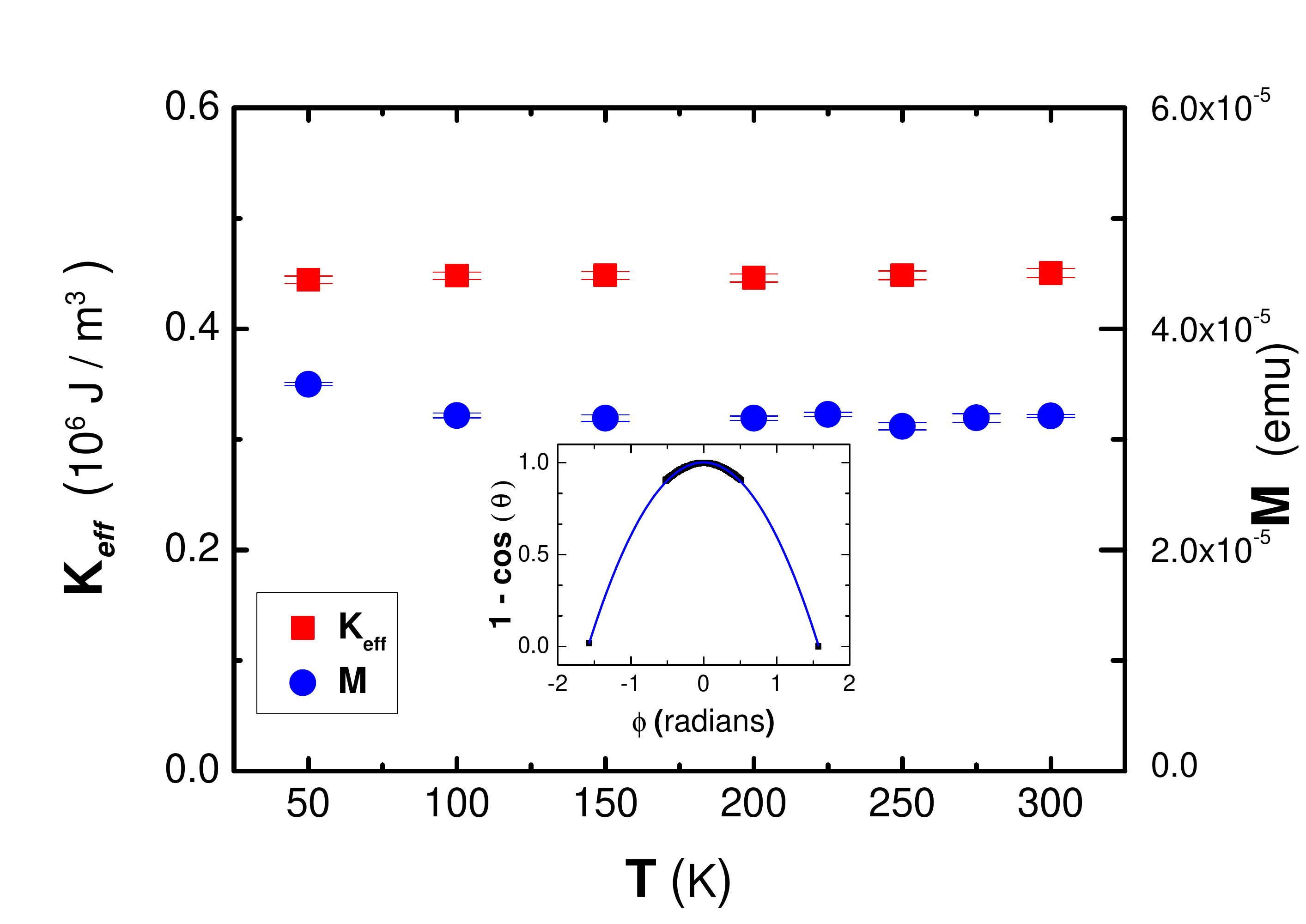}
\caption{Temperature dependence of $K_{eff}$ and $M$.}
\label{fig:Anisoplot}
\end{figure}

{\it Sample characterization}. Firstly, we measure the evolution of the magnetic properties with temperature, since the ferromagnetic layers in these stacks are only a few mono-layers thick. The temperature dependence of the saturation magnetization in this thin film structure is measured using SQUID magnetometry. This is plotted in Fig. \ref{fig:Anisoplot} and in the relevant temperature range a constant value is found.
The temperature dependence of the effective anisotropy is studied by measuring the extraordinary Hall voltage in individual Hall crosses during the application of a rotating magnetic field \cite{moon_determination_2009}. This angle dependent variation of Hall voltage (seen in inset of Fig. \ref{fig:Anisoplot}) when interpreted in terms of the Stoner-Wolfarth model \cite{Stoner599} allows us to determine the anisotropy field and hence the effective anisotropy. The effective anisotropy constant is plotted in Fig. \ref{fig:Anisoplot} and also shows no significant variation. 
These measurements establish that the multilayer has a Curie temperature much larger than 300K, 
with no significant changes in magnetic properties in the considered temperature range.

{\it Spin-orbit torque measurements}.~Next, we perform spin orbit torque measurements using the second-harmonics (2$\omega$) technique \cite{pi_tilting_2010,garello_symmetry_2013,hayashi_quantitative_2014} on patterned Hall bars. The material stack that is used \cite{Supplementary} is: Pt(4.0)$\mid$Co(0.8,1.3)$\mid$AlO$_{x}$(2.0) (all thicknesses in nm).  

\begin{figure}
%\centering
\includegraphics[width = 1\linewidth]{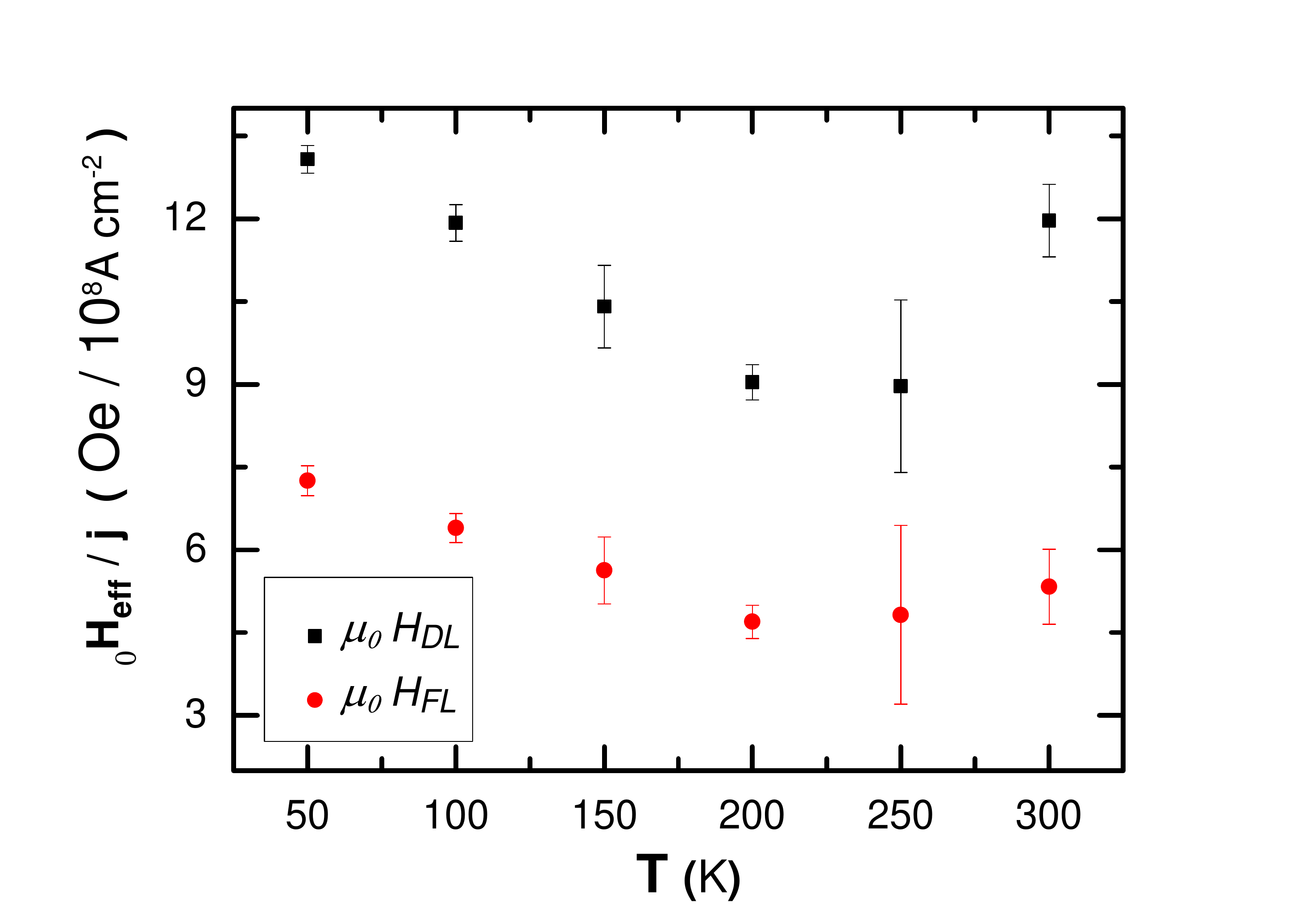}
\caption{Temperature dependence of the efficiency of damping-like ($\mu_0H_{DL}$) and field-like ($\mu_0H_{FL}$) effective fields.}

\label{fig:TempEff}
\end{figure}

The 2$\omega$ measurements \cite{pi_tilting_2010,garello_symmetry_2013,hayashi_quantitative_2014} are performed \cite{Supplementary} by injecting a low frequency ($\omega$/2$\pi$ = 13.7 Hz) sinusoidal ac  signal through the nano wire (x-axis) while the sample is in a saturated magnetization state (z-axis). The effective fields are measured as a function of temperature, and this is shown in Fig. \ref{fig:TempEff}. We see a clear non-monotonic dependence that is distinct from the temperature dependence of the magnetic properties shown in Fig. \ref{fig:Anisoplot}.  

%\begin{figure}[!ht]

The observed temperature dependence of the spin-orbit torques can have various origins, such as: change in magnetic properties, spin-mixing conductance, the SHE and ISGE. We individually evaluate all these parameters to pinpoint the origin of the variation of the spin-orbit torques with temperature. 
First, as already shown, the magnetic properties of our sample are effectively 
temperature independent [see Fig.~\ref{fig:Anisoplot}]. 
Second, experimental results \cite{czeschka_scaling_2011} show that the spin mixing conductance at the Pt$|$Co interface is largely temperature independent as well.  We thus attribute any non-monotonic change in the current-induced effective fields to a variation 
of the SHE and/or ISGE and the resulting torques exerted.

\begin{figure}[!ht]
\centering
\includegraphics[width = 1\linewidth]{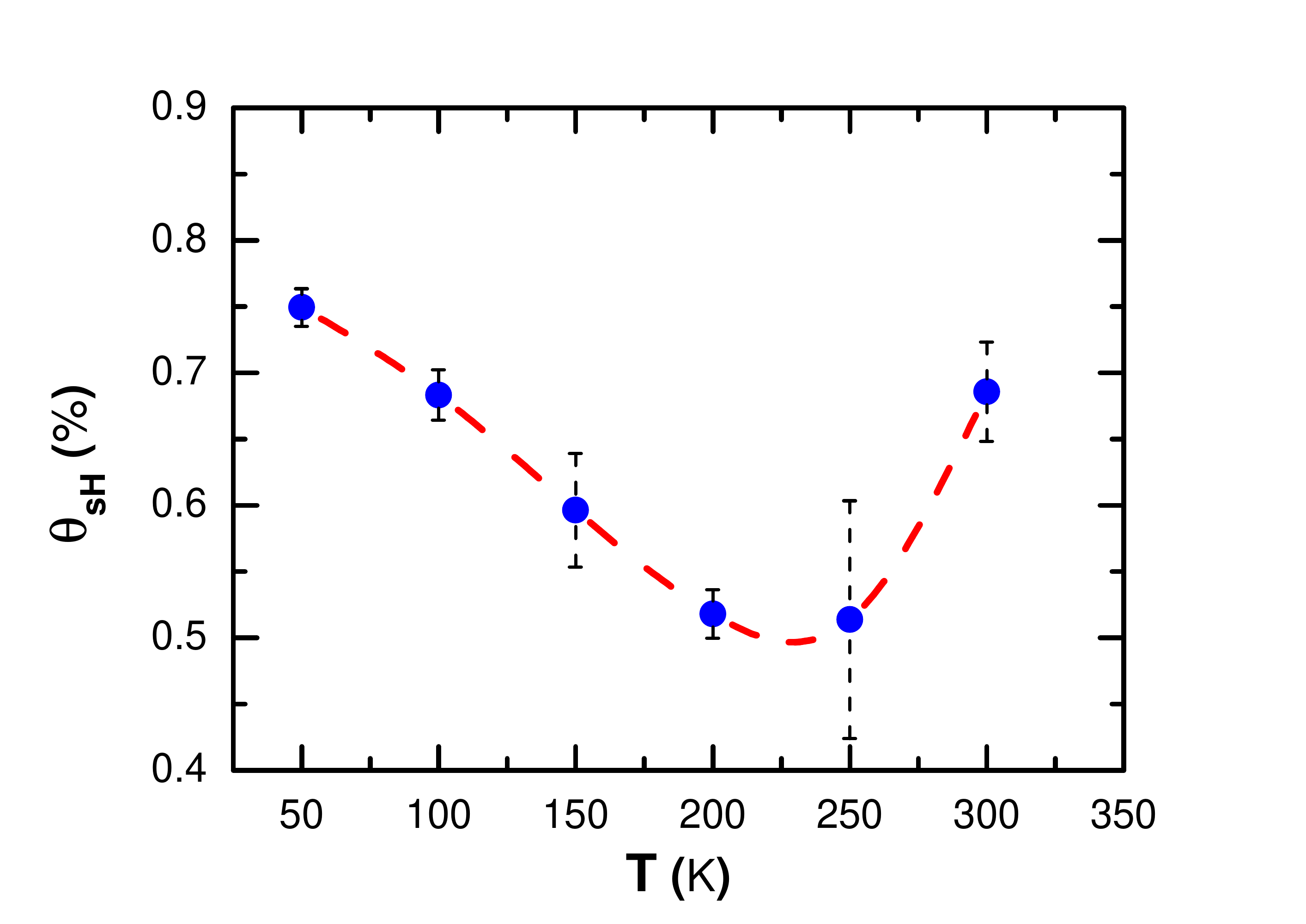}
\caption{Temperature dependence of the deduced spin Hall angle (blue dots $--$ connected by a basis spline as a guide to the eye) derived from the damping-like effective field. Practically identical behavior is found in the control sample, see Fig. S4.}
\label{fig:SHAplot}
\end{figure}

At this point we make a crucial observation, namely that damping- and field-like terms change in unison, 
indicating a common origin for the two torques. 
This agrees with our previous conclusions \cite{conte_ferromagnetic_2016} of the SHE being the likely cause for both SOTs
in Pt$\mid$Co$\mid$AlO$_{x}$, as well as with results where the SHE was claimed to be the dominant effect 
for the damping-like torque in this system \cite{liu_spin-torque_2012,Liu_2012_PRL,qiu_angular_2014}.
As mentioned in the introduction, the generation of both torques via the SHE is due to interfacial 
spin-orbit effects \cite{AminPheno_2016}: 
roughly, part of a $\hat{\sigma}$-polarized spin Hall current $J_{sH}$ incoming on the Co layer is lost and
turned into a $\hat{\sigma}'$-polarized spin current $\delta J_{sH}$ by interfacial spin-orbit coupling.  
Once in the FM, the (remaining) $J_{sH}$ and $\delta J_{sH}$ exert respectively damping-like and field-like torques on the magnetization.  
Since we cannot quantify the loss $\delta J_{sH}$, we define an {\it effective} spin Hall angle
assuming that the remaining spin Hall flux is fully transmitted from Pt to Co
\cite{Seo_2012,Khvalkovskiy_2013,liu_spin-torque_2012,qiu_angular_2014},
\begin{equation}\label{SHA_theta}
	   \theta_{sH} = \frac{2\mu_{0}H_{DL}eM_st_{FM}}
       {\hbar j_e},     
\end{equation}
where $M_s$ is the saturation magnetization and $t_{FM}$ is the ferromagnetic layer thickness and $j_e$ is the injected charge current.
We plot in Fig. \ref{fig:SHAplot} the temperature evolution of the calculated effective spin Hall angle of Pt.

\begin{figure}[!ht]
\centering
\includegraphics[width = 1\linewidth]{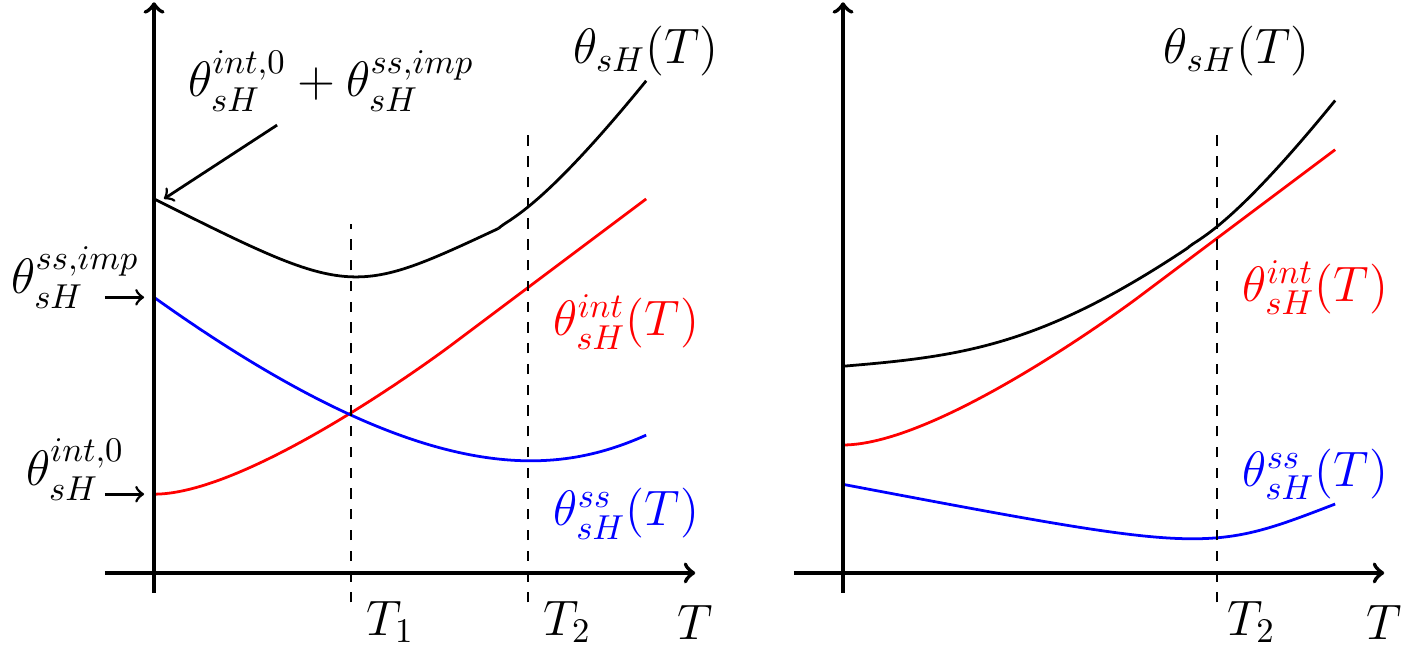}
\caption{
Sketch of the temperature behavior of the SHA for an ultra-clean sample (skew scattering dominated, left panel)
and a dirty one (intrinsic dominated, right panel).
The sketch is based on Eq.~\eqref{new_scaling} and built from knowledge of the high- and low-$T$ asymptotics, 
%respectively $\theta_{sH}^{ss,phon}\sim T$ and $\theta_{sH}^{ss,phon}\rightarrow 0$, 
which imply the existence (absence) of a minimum in clean (dirty) systems\cite{Supplementary}.  
The precise behavior in the intermediate temperature range is however not yet known, nor can accurate values for $T_1, T_2$ be given, since both temperatures
depend on various system parameters.
An order-of-magnitude estimate for Pt yields $T_1 \lesssim T_D \lesssim T_2$, and suggests
that $T_1$ should be higher the cleaner the sample.  In dirty samples the compensation between intrinsic (red)
and skew scattering (blue) curves can give rise to an effectively constant SHA over a broad $T$-range,
as observed in Ta (details see Fig. S4).}
\label{fig:Cosimo}
\end{figure}

{\it Theoretical analysis and discussion}.~We see in Fig. \ref{fig:SHAplot} that the effective SHA varies non-monotonically with temperature. We also observe that there is a significant upturn in its magnitude around 240 - 250 K. This is a distinct temperature in this system, as the spin Hall source in our system, Pt is known \cite{kittel_introduction_2005} to have a Debye temperature of $T_D = 240$ K. The Debye temperature $T_D$ gives 
an estimate of the temperature beyond which all phonon modes are excited and behave classically.

The temperature dependence of the SHA provides us with the opportunity to gain insight into the relative magnitude of the different mechanisms responsible for the SHE in the system. We will start by considering both intrinsic and extrinsic (skew scattering) mechanisms: The importance of the former in 4d and 5d transition metals is known \cite{tanaka_intrinsic_2008}, and so is the fact that in low-impurity samples the dominant extrinsic contribution comes from skew scattering \cite{vignale_ten_2010, Fert_2011,Lowitzer_2011,Hoffmann_2013,Sinova_2015,isasa_temperature_2015}.

We now analyze the temperature dependence of the SHA \cite{isasa_temperature_2015,sagasta_tuning_2016} 
based on the scaling suggested by Tian et al. \cite{Tian_2009, Supplementary}.
The starting point is
\begin{equation}
\label{old_scaling_1}
\rho_{sH} 
\approx
-\left( \sigma_{sH}^{int} + \sigma_{sH}^{ss} \right) \rho^2
\approx
-\sigma_{sH}^{int} \rho^2 + \rho_{sH}^{ss},
\end{equation}
having assumed that skew scattering $(\sigma_{sH}^{ss})$ and intrinsic $(\sigma_{sH}^{int})$ contributions act as parallel channels,
$\sigma_{sH}=\sigma_{sH}^{int} + \sigma_{sH}^{ss}$, and that $\sigma \gg \sigma_{sH}$.  The latter condition implies 
$\sigma_{sH}\approx\rho_{sH}/\rho^2$, with $\rho$ the longitudinal resistance of the Pt layer, and was used in the second passage above.
Then it is argued that the skew scattering resistivity scales as
$ \rho_{sH}^{ss} = \theta_{sH}^{ss, imp} \rho^{imp} + \theta_{sH}^{ss, phon} \rho^{phon} $,
with the spin Hall angle due to phonon skew scattering being temperature {\it independent},
while $\rho^{imp}\;(\rho^{phon})$ represents the resistivity from impurities (phonons).
The split $\rho = \rho^{imp} + \rho^{phon}$ follows Mathiessen's rule.
One thus has
\begin{equation}
\label{old_scaling_2}
\rho_{sH} \approx -\sigma_{sH}^{int} \rho^2 + \theta_{sH}^{ss, imp} \rho^{imp} + \theta_{sH}^{ss, phon} (\rho - \rho^{imp}).
\end{equation}

In the lowest order approximation phonon skew scattering, 
i.e., the last term on the r.h.s. of Eq.~\eqref{old_scaling_2},
is neglected, and $-\rho_{sH}$ can be plotted vs.~$\rho^{2}$.
In this regime, the slope of the linear relation between $\rho_{sH}$ and $\rho^{2}$ 
describes the intrinsic contribution to the SHE, indicating the sign of the SHA of the material.
If this analysis yields inconsistent results, the last term in Eq.~\eqref{old_scaling_2} is also taken into account
\cite{isasa_temperature_2015}.  In our case, however, even the full expression \eqref{old_scaling_2} is not enough, 
since the relation between $\rho_{sH}$ and $\rho$ is non-monotonic, see Fig. S5 and S6.

This suggests the need to revisit the established temperature scaling relation assumed for the SHA.
The microscopic theory of phonon skew scattering shows indeed that the corresponding SHA
is not temperature independent, but rather a function of the temperature, $\theta_{sH}^{ss, phon}\rightarrow\theta_{sH}^{ss, phon}(T)$. 
As a consequence Eq.~\eqref{old_scaling_2} is modified to 
\begin{equation}
\label{new_scaling}
\rho_{sH} \approx -\sigma_{sH}^{int} \rho^2 + \theta_{sH}^{ss, imp} \rho^{imp} + \theta_{sH}^{ss, phon}(T) (\rho-\rho^{imp}).
\end{equation}

The phononic part of the SHA, $\theta_{sH}^{ss,phon}$, increases linearly with $T$ at high temperatures ($T\gtrsim T_D$), while,  
clearly, it has to vanish at low temperatures, though its precise behavior at $T<T_D$ is not yet known
\footnote{The fact that phonon skew scattering arises
from an-harmonic phonon-phonon processes suggests, by analogy with the lattice thermal conductivity, 
a non-monotonic behavior at $T<T_D$.  Notice also that the current phonon skew scattering theory considers only
leading order 3-phonon processes.}.
Its asymptotics can however be used to determine the qualitative behavior of the full SHA in the whole temperature
range.  An order-of-magnitude estimate for a metallic system yields $\sigma^{T_D}_{sH}/\sigma^0_{sH} \lesssim 10^{-1}$,
with $\sigma^{T_D}_{sH} \;(\sigma^0_{sH})$ the spin Hall conductivity at the Debye (zero) temperature \cite{gorini_spin_2015}.
Coupled with the moderate change of the charge conductivity over the same temperature range, $\sigma^{T_D}/\sigma^0 \lesssim 1$, see Fig. S7,
one has $\theta_{sH}^{ss, phon}(T\approx T_D) < \theta_{sH}^{ss, imp} $. 
The resulting qualitative behavior shown in Fig.~\ref{fig:Cosimo} is compatible with the measurements in clean Pt samples.
In a clean system the low-$T$ SHE is dominated by skew scattering,
i.e.~ $-\sigma_{sH}^{int}\rho^2(T\rightarrow0)<\theta_{sH}^{ss, imp} \rho^{imp}$.
Thus, the SHA should eventually decrease in magnitude from $T=0$ down to a minimum at a temperature $T_1$ given by
$-\sigma_{sH}^{int}\rho^2(T_1)\approx\theta_{sH}^{ss, imp} \rho^{imp}$, and increase afterwards.
Furthermore, at a higher temperature $T_2\gtrsim T_D$ such that 
$\theta_{sH}^{ss,phon}(T_2)[\rho(T_2)-\rho^{imp}]\approx\theta_{sH}^{ss,imp}\rho^{imp}$
phonon skew scattering would start dominating impurity skew scattering, yielding an asymptotic
linear behavior:
\begin{equation}
\theta_{sH}(T\gtrsim T_2) \sim 
\left[
\sigma_{sH}^{int} + \sigma_{sH}^{ss, T_D}%(-3\pi N_0 \hbar\Lambda g k_B)
\right] T,
\end{equation}
with $\sigma_{sH}^{ss, T_D}$ the saturation value for the phonon skew scattering spin Hall conductivity at 
$T\gtrsim T_D$ \cite{gorini_spin_2015}.  A few comments are in order: (i) The qualitative analysis yields 
$T_1 \lesssim T_D \lesssim T_2$, but we cannot give quantitative estimates of 
the temperatures $T_1, T_2$.  These require precise knowledge of various system parameters,
notably electron-phonon and phonon-phonon coupling strength.  (ii) The non-monotonic behavior
predicted by Eq.~\eqref{new_scaling} for clean samples does not require phonon skew scattering to be
stronger than impurity skew scattering, which happens only at $T>T_2$: It is enough to recognize
that the phonon SHA is itself a function of the temperature, $\theta_{sH}^{ss, phon}=\theta_{sH}^{ss, phon}(T)$,
vanishing at $T=0$ and (linearly) increasing at high temperatures. (iii) While $\theta_{sH}^{ss,phon}(T\rightarrow0)=0$
has to hold, the approach to zero could be non-trivial and affected e.g.~by the interplay between impurity and phonon scattering
\cite{schmid1973}.

The picture changes in dirty systems with strong intrinsic SHE, such that 
$-\sigma_{sH}^{int}\rho^2(T\rightarrow0)>\theta_{sH}^{ss, imp} \rho^{imp}$.  
In this case Eq.~\eqref{new_scaling} predicts a simpler monotonic behavior of the SHA,
once more compatible with our measurements in a dirty Ta sample (137 $\mu\Omega$cm), see Fig. S4.
Ta has a strong intrinsic SHE \cite{liu_spin-torque_2012}, $T_D$ comparable to that of Pt \cite{Stewart1983},
and our sample is 20 times as resistive as the Pt sample used.
We observe a considerably larger SHA with no minimum, 
indeed roughly constant in the considered temperature range.	

The SHA transition through the intrinsic and extrinsic regimes can be expected due to the low residual resistivity of Pt (5.3 $\mu \Omega$cm, 6.3 $\mu \Omega$cm), as we find \cite{Supplementary} here. This is essentially because the intrinsic and extrinsic contributions scale differently \cite{isasa_temperature_2015} with the resistivity of the metal. 
It was recently shown \cite{sagasta_tuning_2016} that
tuning the conductivity of Pt leads to a transition of the SHE from a ``moderately dirty'' to a ``super-clean'' regime. This corresponds to  a transition from an intrinsic-dominated to an extrinsic-dominated regime. The intrinsic term scales with $\propto $ $\rho^{2}$,
and in the low-resistivity limit is dominated by the phonon contribution, which scales roughly with $\propto $ ($\rho - \rho^{imp}$). 
This could be the reason why other experimental investigations of spin Hall angle of Pt with higher resistivity have observed the dominance of intrinsic contributions, 
while they have observed dominance of extrinsic contribution in case of low resistivity Au \cite{isasa_temperature_2015}.

In summary, we used the second harmonic technique to measure the spin orbit torque in Pt$\mid$Co$\mid$AlO$_{x}$ and Ta$\mid$CoFeB$\mid$MgO. 

We deduced the temperature dependent evolution of the SHA of clean Pt and dirty Ta.
The reproducible non-monotonic behavior observed in Pt cannot be accounted for by standard phenomenology,
but is compatible with the temperature scaling derived from microscopic phonon skew scattering theory.  The monotonic behavior observed in dirty Ta
is instead ascribed to the weakness of skew scattering contributions.
Our results reveal that extrinsic contribution to the SHE might
be the key to understanding the temperature dependence in normal metals as well, and
thus provides a more comprehensive understanding of the
spin Hall angle and the resulting spin orbit torques in this
system.

\begin{acknowledgments}

We acknowledge support by the EU
(MAGWIRE, Project No. FP7-ICT-2009-5; Marie Curie ITN WALL, FP7-PEOPLE-2013-ITN 608031); the DFG (KL1811, TRR 173, TRR 80, SFB 689); Graduate School of Excellence Materials Science in Mainz (MAINZ)
GSC 266; Research Center of Innovative and Emerging Materials
at Johannes Gutenberg University (CINEMA); and the Foundation for Fundamental Research on Matter (FOM), which is part of the Netherlands Organization for Scientific Research (NOW). J.-S. Kim acknowledges supports by the Leading Foreign Research Institute Recruitment Program through the NRF funded by the Ministry of Education, Science and Technology (MEST) (2012K1A4A3053565), the DGIST R$\&$D Program of the Ministry of Science, ICT and Future Planning (17-BT-02 and 17-NT-01).
\end{acknowledgments}

%\bibliography{aipsamp}% Produces the bibliography via BibTeX.

%GVK inserting the BIB file


\end{document}